\documentclass[aps,prb,twocolumn,superscriptaddress,groupedaddress]{revtex4-2}  
\usepackage{amsmath}
\usepackage{graphicx}
\usepackage{subfigure}
\usepackage{color}
\usepackage{amsfonts}
\usepackage{tikz}
\usepackage{esint}
\usepackage{mathrsfs}
\usepackage{hyperref}
\hypersetup{
     colorlinks = true,
     citecolor  = blue,  
     linkcolor  = blue, 
     urlcolor = blue,
     filecolor = blue
}
\begin{document}

\title{Six-component pairing instability in the SU(4) $t$-$J$ chain}

\author{Jia-Cheng He}
\email{jche14@fudan.edu.cn}

\author{Jun-Hao Zhang}

\author{Jie Lou}

\author{Yan Chen}
\email{yanchen99@fudan.edu.cn}
\affiliation{Department of Physics and State Key Laboratory of Surface Physics, Fudan University, Shanghai 200433, China}
\date{\today}

\begin{abstract}
We use the density matrix renormalization group (DMRG) method to study the SU(4) $t$-$J$ chain. We find that, in addition to the conventional repulsive Luttinger liquid phase and phase separation, there are two phases in the attractive Luttinger liquid region dependent on whether the flavor gap is opened or not. The first with the flavor gap is the molecular superfluid phase (the SU(4) singlet instability) which is well-known in the attractive SU(4) Hubbard model ($U<0$). The second without the flavor gap is the superconducting phase (the six-component pairing instability). Furthermore, the molecular superfluid instability cannot coexist with the superconducting instability. This is general in SU($N$) models with $N>2$ and is well demonstrated by the theoretical analysis based on the phenomenological bosonization results.
\end{abstract}


\maketitle

\section{Introduction}
The study on the strong correlations is important in understanding the copper oxide superconductors \cite{Keimer:2015wh}. However, it is very difficult to study the models describing the strongly correlated systems in dimensions bigger than one. Fortunately, the corresponding cases in one dimension have well-developed and extremely successful approaches in both analytical and numerical methods.

Similar to the Fermi liquid theory, Luttinger liquid originating from bosonization technique gives a powerful theoretical framework for one-dimensional systems \cite{book:T.Giamarchi}. All asymptotic properties in a one-dimensional system are completely determined by the unique Luttinger parameter $K$. Even if bosonization is not applicable to a one-dimensional model, one can still obtain the corresponding Luttinger parameter $K$ by numerical methods and thus understand its low-energy properties. The density matrix renormalization group (DMRG) \cite{WhiteDMRGPRL1992,WhiteDMRGPRB1993} as an elegant numerical method is especially effective in dealing with one-dimensional systems. Therefore, DMRG combined with the concepts of Luttinger liquid is a reliable and conventional approach to various one-dimensional systems.
 
The SU(2) $t$-$J$ model is well-known in the investigation of high-temperature superconductivity. Different from the mechanism of the electron-phonon attractive interaction in conventional superconductivity, the $J$ term describing the electron's spin interaction provides an exotic attractive potential for pairing. For example, as witnessed in the numerical study on the one-dimensional SU(2) $t$-$J$ model \cite{PhysRevLett.66.2388,PhysRevB.83.205113}, the existence of the phase separation in values of large $J$ reveals that the $J$ term plays a crucial role in the effective attraction of the system. Note that the one-dimensional Hubbard model has no phase separation. In addition, due to the $J$ term related to spin, one can expect spin orders to compete with superconductivity. In real materials, the orbital degeneracy increasing complexity of the systems is usually found in experiments. Therefore, with orbital degeneracy and spin degeneracy, one can naturally extend the spin symmetry group of the electrons from SU(2) to SU(4) when there is symmetry between the orbital degree and spin degree. Then we can obtain the SU(4) $t$-$J$ model and expect the SU(4) $J$ term to yield the effective attraction. Remarkably, due to the increase in the spin degree, one can expect that in the corresponding superconducting area, more orders yield to join the competition. Moreover, when recovering the orbital index and considering some perturbations breaking SU(4) symmetry, more exotic orders emerge. We believe the SU(4) $J$ term may shed light on some unconventional superconductors. For example, recently, superconductivity near 80 K has been observed in the Ruddlesden-Popper double-layered perovskite nickelate La$_{3}$Ni$_{2}$O$_{7}$ under high pressure \cite{Sun:2023wz}. Under high pressure, this material's Fermi surface yields a small-hole Fermi pocket resulting from lifted 3$d_{z^{2}}$ bonding bands crossing the Fermi level as the apical oxygen ions are hole-doped, and thus the same number of electrons is added to Ni 3$d_{x^{2}-y^{2}}$ orbitals of electron bands.  The electrons of Ni 3$d_{x^{2}-y^{2}}$ interacted with oxygen 2$p$ orbitals are expected to yield the intra-layer Zhang-Rice singlets \cite{PhysRevB.37.3759}. The enlarged splitting of the 3$d_{z^{2}}$ orbitals indicates the strong electronic interactions between the layers of NiO$_{2}$. Due to the double-layered structure, there is degeneracy in orbitals. According to the similarities to copper oxide superconductors, including the phase diagram, we propose the SU(4) $J$ term plays a dominant role in the interaction of the system.

This paper is organized as follows. In Sec. \ref{sec:su4tJchain} the model is introduced. In Sec. \ref{sec:correlation-functions} the correlation functions are listed. In Sec. \ref{sec:pheno-bosonization-analysis} we present the theoretical analysis for SU($N$) models based on the phenomenological bosonization results. In Sec. \ref{sec:structure-factors} the DMRG results of structure factors of various phases in the SU(4) $t$-$J$ chain are presented. In Sec. \ref{sec:SC-area-without-the-flavor-gap} we are devoted to analyzing the DMRG results of the superconducting phase in the SU(4) $t$-$J$ chain to verify the theoretical conclusions based on bosonization. The discussion and conclusion will be given in Sec. \ref{sec:Discussionandconclusion}.

\section{model}\label{sec:su4tJchain}
The one-dimensional SU($N$) Hubbard model reads
\begin{equation}\label{eq: su(4) Hubbard model}
H^{\text{SU}(N)}=-t\sum_{i}\sum_{\alpha=1}^{N}(c^{\dagger}_{i,\alpha}c_{i+1,\alpha}+\mathrm{h.c.})+U\sum_{i}(\sum_{\alpha=1}^{N}\hat{n}_{i,\alpha})^{2},
\end{equation}
where $c_{i,\alpha}$ represents annihilating a fermion on site $i$ with flavor $\alpha$, and $\hat{n}_{i,\alpha}=c^{\dagger}_{i,\alpha}c_{i,\alpha}$. We consider the large $U$ limit case of SU($N$) Hubbard model and thus obtain the one-dimensional SU($N$) $t$-$J$ model:
\begin{flalign}\label{eq:SUNt-Jmodelv1}
H_{t-J}^{\text{SU}(N)}=&-t\sum_{i\alpha}P_{G}(c_{i,\alpha}^{\dagger}c_{i+1,\alpha}+c_{i+1,\alpha}^{\dagger}c_{i,\alpha})P_{G}
                    \nonumber\\
&+J\sum_{i\alpha\alpha^{\prime}}(c_{i,\alpha}^{\dagger}c_{i,\alpha^{\prime}}c_{i+1,\alpha^{\prime}}^{\dagger}c_{i+1,\alpha}-\frac{1}{\nu_{0}}\hat{n}_{i,\alpha}\hat{n}_{i+1,\alpha^{\prime}}),
\end{flalign}
where $J=t^{2}/U$ and $\nu_{0}=1,2,\dots,N-1$. This Hamiltonian represents the case of hole doping away from the integer filling number $\nu_{0}$. The projection operator $P_{G}$ represents excluding the particle occupancy states which stay at very high-energy levels arising from the Hubbard U term, and only permits the particle number on each site to be $\nu_{0}$ or $\nu_{0}-1$.
We can consider the value of $J$ is independent of $t$ in the $t$-$J$ model. However, the $t$-$J$ model can no longer be associated with a Hamiltonian with a density-density interaction in the case of large $J$. Note that the case of $J=t$ with $\nu_{0}=1$ corresponds to the supersymmetric limit of the SU($N$) $t$-$J$ model \cite{Schlottmann_1993}. In this paper, we only study the SU(4) $t$-$J$ chain in the case of $\nu_{0}=1$:
\begin{flalign}\label{eq:standardSU4tJmodelv01}
H_{t-J}^{\text{SU}(4)}=&-t\sum_{i\alpha}P_{G}(c_{i,\alpha}^{\dagger}c_{i+1,\alpha}+c_{i+1,\alpha}^{\dagger}c_{i,\alpha})P_{G}
                  \nonumber\\
&+J\sum_{i\alpha\alpha^{\prime}}(c_{i,\alpha}^{\dagger}c_{i,\alpha^{\prime}}c_{i+1,\alpha^{\prime}}^{\dagger}c_{i+1,\alpha}-\hat{n}_{i,\alpha}\hat{n}_{i+1,\alpha^{\prime}}).
\end{flalign}
We set $t=1$. We used ITensor library \cite{itensor} for numerical calculations. In our DMRG numerical calculations, the states can be kept up to $m=5000$,  the number of sweeping is enough to obtain the convergent data, and the truncation errors are smaller than $10^{-6}$. We used the open boundary condition and the size of the system as $L=80$ sites.

\section{correlation functions}\label{sec:correlation-functions}
We need to measure some correlation functions to discriminate these different phases of the SU(4) $t$-$J$ chain. We set the lattice constant $a_{0}=1$. The density-density correlation function and its structure factor read
\begin{flalign}
&N_{ij}=\langle \hat{n}_{i}\hat{n}_{j}\rangle-\langle \hat{n}_{i}\rangle\langle \hat{n}_{j}\rangle,
                         \nonumber\\
&N(k)=\frac{1}{L}\sum_{i,j=1}^{L}N_{ij}e^{ik(x_{i}-x_{j})},
\end{flalign}
where $\hat{n}_{i}=\sum_{\alpha=1}^{4}\hat{n}_{i,\alpha}$. The flavor-flavor correlation function and its structure factor read
\begin{flalign}\label{eq:T23correlation}
&T^{2(3)}_{i j}=\left\langle\hat{T}_{i}^{2(3)} \hat{T}_{j}^{2(3)}\right\rangle,
                  \nonumber\\
&T^{23}(k)=\frac{1}{L}\sum_{i,j=1}^{L}T^{2(3)}_{ij}e^{ik(x_{i}-x_{j})},
\end{flalign}
where $\hat{T}_{i}^{2(3)}=\sqrt{2}\left(c_{i 1}^{\dagger} c_{i 1}-c_{i 2}^{\dagger} c_{i 2}\right)$. Similar to the SU(4) Hubbard model, the SU(4) $t$-$J$ model has three flavor degrees of freedom. Due to SU(4) symmetry, calculating one of them is enough. One-particle density matrix and momentum distribution function read
\begin{flalign}
&n^{\alpha\alpha}_{ij}=\langle c_{i,\alpha}^{\dagger}c_{j,\alpha}\rangle,
             \nonumber\\       
&n_{\alpha}(k)=\frac{1}{L}\sum_{i,j=1}^{L}n^{\alpha\alpha}_{ij}e^{ik(x_{i}-x_{j})},
\end{flalign}
and note that $n_{ij}^{\alpha\beta}=0$ for $\alpha\ne\beta$. Due to SU(4) symmetry, it is enough to only consider $n^{\alpha\alpha}$ in one flavor. Therefore, we denote $n_{\alpha}(k)$ as $n(k)$. The SU(4) singlet state can be expressed as
\begin{flalign}
\mathcal{M}_{i}^{\dagger}=\frac{1}{\sqrt{4!}}\sum_{\alpha,\beta,\gamma,\delta}\epsilon_{\alpha\beta\gamma\delta} c_{i,\alpha}^{\dagger} c_{i+1,\beta}^{\dagger}c_{i+2,\gamma}^{\dagger}c_{i+3,\delta}^{\dagger},
\end{flalign}
where $\epsilon_{\alpha\beta\gamma\delta}$ is forth order antisymmetric tensor. Its correlation function is
\begin{flalign}
Q_{ij}=\langle \mathcal{M}_{i}^{\dagger}\mathcal{M}_{j}\rangle.
\end{flalign}
The interaction term of the SU($4$) $t$-$J$ model in Eq. (\ref{eq:standardSU4tJmodelv01}) can be rewritten in the representation of the SU($4$) generators. More details can be found in Appendix \ref{sec:time-dependent-correl}. Then, by using the Fierz identity, the interaction term can be rewritten as \cite{CenkeXu,PhysRevB.105.245117}
\begin{equation}\label{eq:pairingform}
\sum_{a=1}^{15} \hat{T}_{i}^{a} \hat{T}_{j}^{a}=-\frac{5}{4}\left(\overrightarrow{\Delta}_{i j}\right)^{\dagger} \cdot \overrightarrow{\Delta}_{i j}+\frac{3}{4}\left(\Delta_{i
j}^{-}\right)^{\dagger} \cdot \Delta_{i j}^{-},
\end{equation}
where $\overrightarrow{\Delta}_{i j}$ and $\Delta_{i j}^{-}$ are the pairing fields with 6 and 10 components, respectively. The energy of the pairing field $\overrightarrow{\Delta}_{i j}$ is lower than that of $\Delta_{i j}^{-}$ due to $J>0$. By choosing a representation, these two pairing fields can be expressed as:
\begin{flalign}
&\overrightarrow{\Delta}_{i j}=\sum_{m=1}^{6}\left(\sum_{\alpha\beta}c_{i\alpha}\Gamma_{\alpha\beta}^{n_{m}}c_{j\beta}\right)\hat{\mathbf{e}}_{m},
                        \nonumber\\
&\Delta_{i j}^{-}=\sum_{m=1}^{10}\left(\sum_{\alpha\beta}c_{i\alpha}\Gamma_{\alpha\beta}^{p_{m}}c_{j\beta}\right)\hat{\mathbf{e}}_{m},
\end{flalign}
where $\Gamma$ is the generator of SU(4) group satisfying the relation $\mathrm{Tr}(\Gamma^{a}\Gamma^{b})=4\delta_{ab}$. And we use the superscript $n_{m}$ to denote the antisymmetric six-component generators of SU(4) group, and the superscript $p_{m}$ to denote the nine-component generators of SU(4) group including the symmetric elements and the identity element. Here we use $\hat{\mathbf{e}}_{m}$ to represent an orthogonal normalized vector basis. We only consider the lower-level pairing field $\overrightarrow{\Delta}_{i j}$. In addition, it is enough to only consider one of the six components of $\overrightarrow{\Delta}_{i j}$ due to the symmetry between them, such as
\begin{flalign}
\Delta_{i}^{s\dagger}=\left(c_{i, 1}^{\dagger} c_{i+1,2}^{\dagger}-c_{i, 2}^{\dagger} c_{i+1,1}^{\dagger}\right) / \sqrt{2}.
\end{flalign}
The corresponding correlation function reads
\begin{flalign}
P_{s,ij}=\langle \Delta_{i}^{s\dagger}\Delta_{j}^{s}\rangle.
\end{flalign}

\section{phenomenological bosonization results for the correlation functions of SU($N$) models}\label{sec:pheno-bosonization-analysis}
The projection operator prevents the $t$-$J$ model from being solved analytically except in the case of supersymmetric limit \cite{Schlottmann_1993}. Therefore, we resort to the phenomenological bosonization. In this section, we give the phenomenological bosonization results of SU($N$) models at zero temperature for the convenience of subsequent discussions.

Bosonization illustrates that for SU($N$) models at low energy the charge and flavor degrees of freedom are separated into sectors all described by the Luttinger liquids \cite{book:T.Giamarchi,PhysRevB.60.2299}. The charge sector depends on the Luttinger parameter $K_{\rho}$. The $N-1$ flavor sectors depend on the Luttinger parameter $K_{\sigma}$ under SU($N$) symmetry. In the following, we list various equal-time correlation functions for the case of the gapless regime with gapless modes both in the charge sector and the flavor sectors. The total density correlation function is given by
\begin{flalign}\label{eq:pheno-boson-totdensity}
&\langle \rho(x)\rho(0) \rangle=n^{2}-\frac{NK_{\rho}}{2(\pi x)^{2}}
                                       \nonumber\\
&+\sum_{p=1}^{\infty}A_{p+1}\cos(2pk_{F}x)\left(\frac{\alpha}{|x|}\right)^{2p^{2}[K_{\rho}/N+(1-1/N)K_{\sigma}]}
                                       \nonumber\\
&=n^{2}-\frac{NK_{\rho}}{2(\pi x)^{2}}+A_{2}\cos(2k_{F}x)\left(\frac{\alpha}{|x|}\right)^{2[K_{\rho}/N+(1-1/N)K_{\sigma}]}
                                       \nonumber\\
&+\cdots,
\end{flalign}
where $k_{F}=\pi n/N$. $n$ is the average density of particles, and $\alpha$ is a cutoff. The flavor-flavor correlation function is given by
\begin{flalign}\label{eq:pheno-boson-flavorcorrel}
&\langle [\rho_{1}(x)-\rho_{2}(x)][\rho_{1}(0)-\rho_{2}(0)]\rangle
=-\frac{K_{\sigma}}{(\pi x)^{2}}
                 \nonumber\\
&+\sum_{p=1}^{\infty} B_{p+1} \cos(2pk_{F}x)\left(\frac{\alpha}{|x|}\right)^{2p^{2}[K_{\rho}/N+(1-1/N)K_{\sigma}]}
                 \nonumber\\
&=-\frac{K_{\sigma}}{(\pi x)^{2}}+B_{2} \cos(2k_{F}x)\left(\frac{\alpha}{|x|}\right)^{2[K_{\rho}/N+(1-1/N)K_{\sigma}]}
                 \nonumber\\
&+\cdots.
\end{flalign}
Note that, in Eqs. (\ref{eq:pheno-boson-totdensity}) and (\ref{eq:pheno-boson-flavorcorrel}), we don't consider the corrections from the interaction of flavor sectors which possibly eliminate the exponent $K_{\sigma}$ or reduce it in the $4k_{F}$, $6k_{F}$, and $8k_{F}$ terms etc. except for the $2k_{F}$ term. We also don't consider the logarithmic corrections for the $2k_{F}$ term in Eqs. (\ref{eq:pheno-boson-totdensity}) and (\ref{eq:pheno-boson-flavorcorrel}). The correlation function of the SU($N$) singlet state is given by
\begin{flalign}\label{eq:SU(N)singletcorrelbosoneven}
\langle\widetilde{\mathcal{M}}^{\dagger}(x)\widetilde{\mathcal{M}}(0)\rangle&=\sum_{p=0}^{\infty}C_{p}\left(\frac{\alpha}{|x|}\right)^{(2p+1)^{2}NK_{\sigma}/2+NK_{\rho}^{-1}/2}
              \nonumber\\
&+\cdots
              \nonumber\\
&=C_{0}\left(\frac{\alpha}{|x|}\right)^{NK_{\sigma}/2+NK_{\rho}^{-1}/2}+\cdots,
\end{flalign}
where $\widetilde{\mathcal{M}}^{\dagger}(x)=c^{\dagger}_{x,1}c^{\dagger}_{x,2}\cdots c^{\dagger}_{x,N}$. Note that $N$ in Eq. (\ref{eq:SU(N)singletcorrelbosoneven}) is restricted to even numbers. The case of odd numbers for $N$ is given by
\begin{flalign}\label{eq:SU(N)singletcorrelbosonodd}
&\langle\widetilde{\mathcal{M}}^{\dagger}(x)\widetilde{\mathcal{M}}(0)\rangle=\sum_{p=0}^{\infty}C^{\prime}_{p}\sin[(2p+1)k_{F}x]
               \nonumber\\
&\left(\frac{\alpha}{|x|}\right)^{(2p+1)^{2}[K_{\rho}/(2N)+(N-1/N)K_{\sigma}/2]+NK_{\rho}^{-1}/2}+\cdots
              \nonumber\\
&=C^{\prime}_{0}\sin(k_{F}x)\left(\frac{\alpha}{|x|}\right)^{K_{\rho}/(2N)+NK_{\rho}^{-1}/2+(N-1/N)K_{\sigma}/2}
              \nonumber\\
&+\cdots.
\end{flalign}
In addition, when $K_{\sigma}=1$, Eqs. (\ref{eq:pheno-boson-totdensity}), (\ref{eq:pheno-boson-flavorcorrel}), (\ref{eq:SU(N)singletcorrelbosoneven}) and (\ref{eq:SU(N)singletcorrelbosonodd}) recover the case of the SU($N$) Hubbard model \cite{Capponi:2016uy,PhysRevA.84.043601,PhysRevA.77.013624}. The correlation function of the flavor-antisymmetric pairing is given by
\begin{flalign}\label{eq:antisym-pairing-correl}
&\langle\Delta^{s\dagger}(x)\Delta^{s}(0)\rangle
                                \nonumber\\
&=\sum_{p=0}^{\infty}D_{p}\left(\frac{\alpha}{|x|}\right)^{2/(NK_{\rho})+(1-2/N)K_{\sigma}^{-1}+(2p+1)^{2}K_{\sigma}}+\cdots
                                \nonumber\\
&=D_{0}\left(\frac{\alpha}{|x|}\right)^{2/(NK_{\rho})+(1-2/N)K_{\sigma}^{-1}+K_{\sigma}}+\cdots,
\end{flalign}
Note that the amplitudes $A_{p+1}$, $B_{p+1}$, $C_{p}$, $C^{\prime}_{p}$ and $D_{p}$ are non-universal coefficients. 

If one correlation function $R(r)$ decay as a power law $R(r)\sim e^{i2k_{F}x}r^{-\upsilon}$ (here $r=\sqrt{x^{2}+(u\tau)^{2}}$, where $u$ is a velocity and $\tau$ is the imaginary time.), then at zero temperature, the corresponding susceptibility is given by
\begin{flalign}\label{eq:susceptibility-nu2}
\chi(k,\omega)\sim\mathrm{max}[\delta k,\omega]^{\upsilon-2},
\end{flalign}
where $\omega$ is the frequency. This relation indicates that the susceptibility diverges for $\upsilon<2$. Although there is no true order in the one-dimensional system according to Mermin-Wagner theorem \cite{PhysRevLett.17.1133}, a divergent susceptibility means the system would like to order into a state. The time-dependent form of correlation functions Eqs. (\ref{eq:pheno-boson-totdensity}), (\ref{eq:pheno-boson-flavorcorrel}), (\ref{eq:SU(N)singletcorrelbosoneven}) and (\ref{eq:antisym-pairing-correl}) can be found in Appendix \ref{sec:time-dependent-correl}, and note that the order of the time in correlation functions is the same with that of the distance.

Since the system has the flavor rotation invariance, the $K_{\sigma}=1$ remains unchanged in the Luttinger liquid area of flavor sectors. Therefore, according to Eq. (\ref{eq:susceptibility-nu2}), we can easily distinguish the different phases in the gapless regime with gapless modes both in the charge sector and the flavor sectors ($K_{\sigma}=1$) by calculating the Luttinger parameter $K_{\rho}$. According to the leading terms in Eqs. (\ref{eq:pheno-boson-totdensity}) and (\ref{eq:antisym-pairing-correl}), $K_{\rho}<1$ represents the repulsive liquid. $K_{\rho}=1$ represents the free system. $K_{\rho}>1$ represents the phase of attractive interaction. $K_{\rho}$ can be obtained by calculating the slope of the structure factor of the density-density correlation function when $k\to0$. By Fourier transformation of Eq. (\ref{eq:pheno-boson-totdensity}), the relation reads \cite{PhysRevA.84.043601}
\begin{equation}\label{eq:SUNNkKrho}
N(k \rightarrow 0)=\frac{N K_{\rho}}{2 \pi}|k|.
\end{equation}
Note that Eq. (\ref{eq:SUNNkKrho}) is only suitable for the gapless regime in the charge sector. For the gapped regime in the charge sector or the flavor sector, in the long wavelength range, we have a relation 
\begin{flalign}\label{eq:gapedSUNNk}
N_{\nu}(k)=\frac{C_{\nu}u_{\nu}K_{\nu}}{2\pi}\frac{k^{2}}{\sqrt{(u_{\nu}k)^{2}+\Delta^{2}}},
\end{flalign}
where $\Delta$ is the gap, and $\nu=\rho$ or $\sigma$ denotes the charge sector or the flavor sector, respectively. $C_{\rho}=N$, and $C_{\sigma}=2$. $u_{\nu}$ is the characteristic velocity of the excitation in the corresponding sector. In addition, $N_{\sigma}(k)=T^{23}(k)/2$ in the long wavelength range. We see that when $\Delta=0$ the Eq. (\ref{eq:gapedSUNNk}) recovers Eq. (\ref{eq:SUNNkKrho}). When $\Delta\ne 0$, we obtain $N_{\nu}(k\rightarrow 0)\sim k^{2}$. The quadratic behavior can be clearly distinguished from the linear behavior. Therefore, we can discriminate between the gapped regime and the gapless regime according to this feature. 

The correlation functions for the gapped regime in the charge sector or the flavor sector can be obtained directly by letting $K_{\nu}\rightarrow 0$ in Eqs. (\ref{eq:pheno-boson-totdensity}), (\ref{eq:pheno-boson-flavorcorrel}), (\ref{eq:SU(N)singletcorrelbosoneven}), (\ref{eq:SU(N)singletcorrelbosonodd}) and (\ref{eq:antisym-pairing-correl}). One important case is that when $K_{\rho}>1$ and $N>2$ the results of whether the flavor gap is opened or not are completely opposite. When the flavor gap is opened ($K_{\sigma}\rightarrow 0$), the exponent $(1-2/N)K^{-1}_{\sigma}\rightarrow \infty$ of the leading terms in Eq. (\ref{eq:antisym-pairing-correl}) indicates the correlation function of the pairing is exponentially suppressed and thus its susceptibility is not divergent any more. In contrast, the correlation of the SU($N$) singlet is strongly enhanced according to Eqs. (\ref{eq:SU(N)singletcorrelbosoneven}) and (\ref{eq:SU(N)singletcorrelbosonodd}). This is easily observed when $N=4$, the exponent $2K^{-1}_{\rho}$ of the leading terms in Eq. (\ref{eq:SU(N)singletcorrelbosoneven}) indicates its susceptibility is divergent. The superfluid consisting of SU($N$) singlets with $N>2$ is called a molecular superfluid \cite{Capponi:2016uy}. This case is well-known in the attractive SU($N$) Hubbard model ($U<0$) \cite{Capponi:2016uy,PhysRevA.77.013624,Roux:2009um}. However, when the flavor gap is not opened ($K_{\sigma}=1$), the exponent of the leading term in the correlation function of the SU($N$) singlet (with $N>2$) is bigger than two according to Eq. (\ref{eq:SU(N)singletcorrelbosoneven}) or Eq. (\ref{eq:SU(N)singletcorrelbosonodd}), indicating its susceptibility is not divergent any more. For example, in the case of $N=4$, its exponent of the leading term is given by $2+2K_{\rho}^{-1}>2$. In contrast, the exponent $2(K_{\rho}^{-1}-1)/N+2<2$ of the leading term in the correlation function of the pairing indicates its susceptibility is divergent according to Eq. (\ref{eq:antisym-pairing-correl}). Therefore, the molecular superfluid instability (the SU($N$) singlet instability) cannot coexist with the superconducting instability (the pairing instability) in the case of $N>2$ and $K_{\rho}>1$. In addition, the molecular superfluid instability may occur only when the flavor gap is opened (in the SU(4) case, the additional condition for the molecular superfluid instability is $K_{\rho}>1$), and the condition for the superconducting instability is $K_{\rho}>1$ with the gapless modes in the flavor sectors, according to their correlation functions listed above in SU($N$) models with $N>2$.

\begin{figure}
\centering
\includegraphics[width=0.48\textwidth]{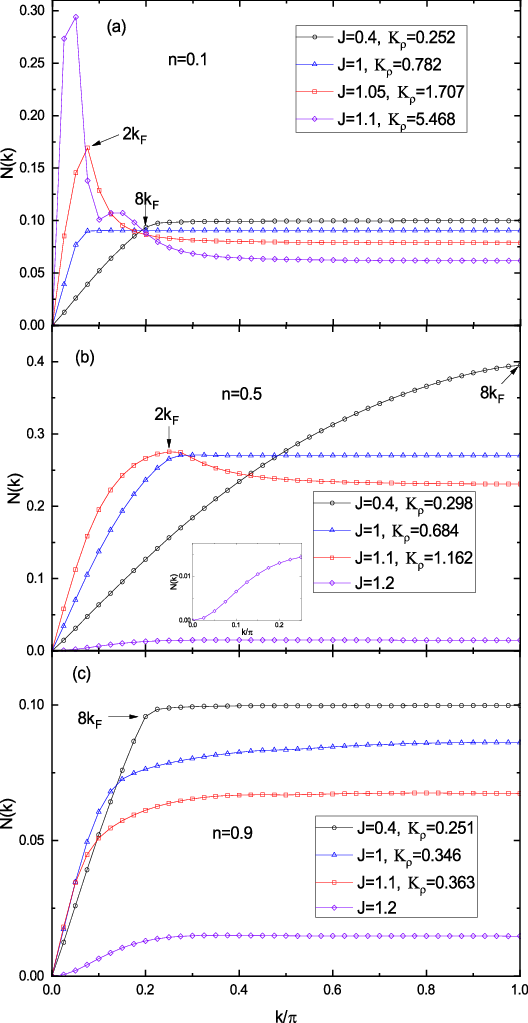}
\caption{\label{normal_multi_Nk_rightHL80_su4tJchain}The structure factor $N(k)$ of the density-density correlation function with various values of $n$ and various values of $J$. The size of the system is given by $L=80$ sites. (a) The case of $n=0.1$. (b) The case of $n=0.5$. (Inset) The enlargement of the case of $J=1.2$ near small $k$, which presents a quadratic behavior ($\sim k^{2}$). (c) The case of $n=0.9$, here the $8k_{F}$ resulting from the one folded back to the sector of the first Brillouin zone with $k>0$.}
\end{figure}

\begin{figure}
\centering
\includegraphics[width=0.48\textwidth]{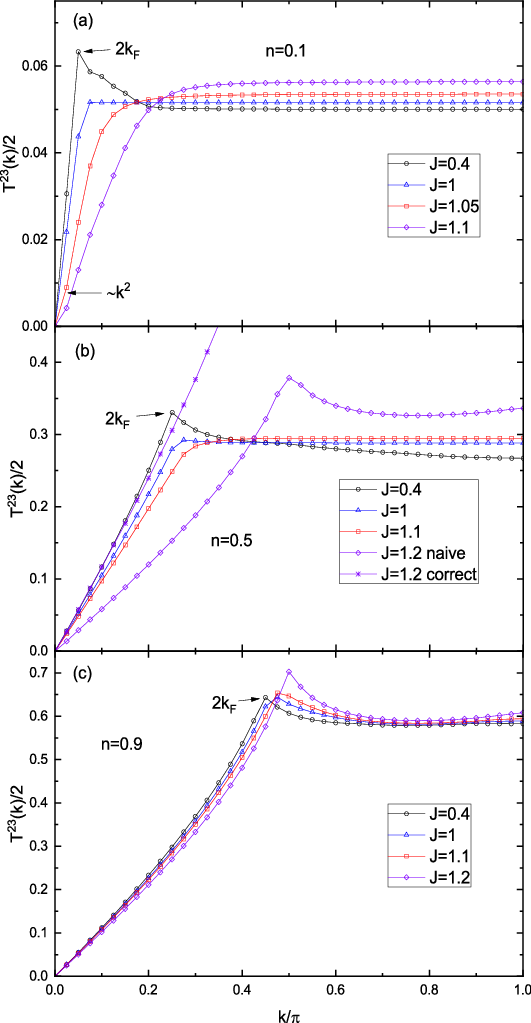}
\caption{\label{normal_multi_Sk_rightHL80_su4tJchain}The structure factor $T^{23}(k)$ of the flavor-flavor correlation function with various values of $n$ and various values of $J$. The size of the system is given by $L=80$ sites. (a) The case of $n=0.1$. (b) The case of $n=0.5$. Note that the data of ``$J=1.2$ correct" corresponds to the case of the effective size of the system, $L^{\mathrm{effective}}=L/2=40$ sites, due to the frozen charge degree of freedom. (c) The case of $n=0.9$. }
\end{figure}

\section{DMRG results: structure factors}\label{sec:structure-factors}
Subsequently, we present the numerical results of various structure factors of systems with various values of the average density of particles $n=N_{e}/L$ ($N_{e}$ is the total number of particles) and various values of $J$.

\begin{figure}
\centering
\includegraphics[width=0.48\textwidth]{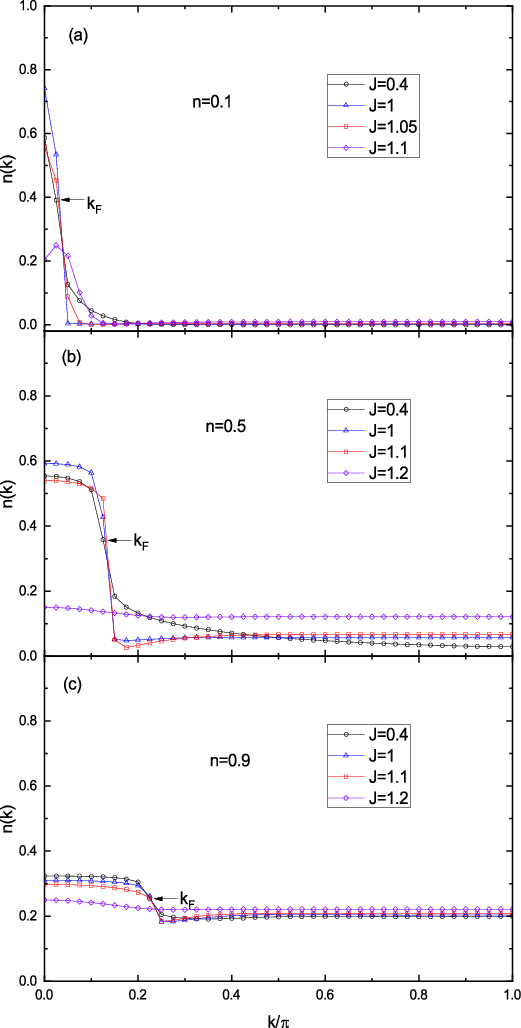}
\caption{\label{normal_multi_nnk_rightHL80_su4tJchain}The momentum distribution function $n(k)$ with various values of $n$ and various values of $J$. The system size is given by $L=80$ sites. (a) The case of $n=0.1$. (b) The case of $n=0.5$. (c) The case of $n=0.9$.}
\end{figure}

Figure \ref{normal_multi_Nk_rightHL80_su4tJchain} shows the structure factor $N(k)$ of the density correlation function. The cases of $J=0.4$ with all values of $n$ have the common properties including $K_{\sigma}=1$, $K_{\rho}<1$ and the $8k_{F}$ ($k_{F}=n\pi/4$) anomaly. As said above, $K_{\rho}<1$ indicates that the system stays in the repulsive Luttinger liquid phase. The anomaly $8k_{F}=2\pi n$ illustrates a repulsive interaction between particles. Their structure factors $T^{23}(k)$ of the flavor correlation function all have a $2k_{F}$ peak, as shown in Fig. \ref{normal_multi_Sk_rightHL80_su4tJchain}. This feature combined with the $8k_{F}$ anomaly of $N(k)$ indicates there is an instability towards the flavor density wave rather than the charge density wave (CDW). In addition, their momentum distribution functions all possess the feature of power-law decay near the Fermi surface, as shown in Fig. \ref{normal_multi_nnk_rightHL80_su4tJchain}.

When the value of $J$ increases to 1, corresponding to the supersymmetric limit, the $8k_{F}$ anomaly in $N(k)$ disappears in all values of $n$, and the $2k_{F}$ peak of $T^{23}(k)$ also almost disappears. 

With the value of $J$ increasing into the range of $1<J<1.2$, we observe that a $2k_{F}$ peak of the $N(k)$ appears in low and intermediate densities, as shown in Figs. \ref{normal_multi_Nk_rightHL80_su4tJchain}(a) and \ref{normal_multi_Nk_rightHL80_su4tJchain}(b). This $2k_{F}$ peak indicates the CDW instability yielded in the system. Note that the value of $k$ corresponding to the position of the $2k_{F}$ peak in Fig. \ref{normal_multi_Nk_rightHL80_su4tJchain}(a) is slightly larger than the theoretical value $n\pi/2$, which results from the open boundary condition we adopted. The open boundary condition imposes that the electron density is almost zero near the boundary, which in return reduces the effective number of lattice sites and thus increases the effective particle density. This discrepancy is most evident in the low-density area and reduces in the high-density area, for example, this discrepancy almost disappears in the case of $n=0.5$ as shown in Fig. \ref{normal_multi_Nk_rightHL80_su4tJchain}(b). Their $K_{\rho}>1$ indicates there is a superconducting instability or a molecular superfluid instability in this area according to the theoretical analysis in Sec. \ref{sec:pheno-bosonization-analysis}. Their $T^{23}(k)$ does not have the $2k_{F}$ peak, as shown in Figs. \ref{normal_multi_Sk_rightHL80_su4tJchain}(a) and \ref{normal_multi_Sk_rightHL80_su4tJchain}(b). In the cases of $n=0.1$ with $J=1.05$ and $J=1.1$, we can clearly observe that $T^{23}(k)/2$ possesses the quadratic behavior near small $k$, as shown in Fig. \ref{normal_multi_Sk_rightHL80_su4tJchain}(a). This feature indicates the existence of the flavor gap according to Eq.
(\ref{eq:gapedSUNNk}). The phase of the gapless charge mode with the flavor gap is also referred to as the Luther-Emery liquid \cite{PhysRevLett.33.589}. The existence of the flavor gap with $K_{\rho}>1$ indicates there is a molecular superfluid instability in the case of $n=0.1$ with $J=1.05$. In this case, the exponent $2/K_{\rho}=1.1716$ of the leading term in the correlation function of the SU(4) singlet is evidently bigger than the exponent $K_{\rho}/2=0.8535$ of the leading term in the density correlation function. This indicates the CDW instability is more dominant than the molecular superfluid instability. The case of $n=0.1$ with $J=1.1$ will be discussed later due to its difference and complexity. In the case of $n=0.5$ with $J=1.1$, its $T^{23}(k)/2$ keeps linear behavior near small $k$, indicating gapless modes in flavor sectors and thus a superconducting instability. As shown in Fig. \ref{normal_multi_nnk_rightHL80_su4tJchain}(a), in the case of $n=0.1$ and $J=1.05$, the weight of the dispersion evidently reduces near the outside of the Fermi surface and increases in the area of $k>k_{F}$ with $k$ far from $k_{F}$ compared to that of the repulsive Luttinger liquid phase. The latter feature is not very evident in the low-density area but in the higher-density area, for example, in the case of $n=0.5$ with $J=1.1$.

\begin{figure*}
\centering
\includegraphics[width=1\textwidth]{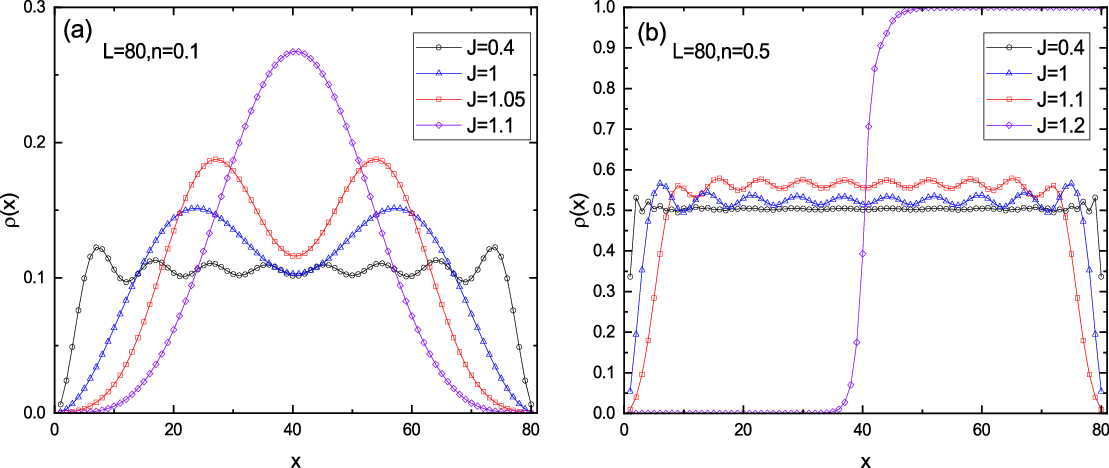}
\caption{\label{normal_leftright_multi_rhox_rightHL80}The particle density distribution in real space $\rho(x)$. The system size is given by $L=80$ sites. (a) In the case of $n=0.1$, due to the open boundary condition we can clearly observe the Friedel oscillation, in the repulsive Luttinger liquid area with $J=0.4$, with each particle yielding a wave packet. Increasing the value of $J$ to the case of the supersymmetry limit ($J=1$) or the case of the molecular superfluid instability ($J=1.05$), every four particles yield a wave packet. Further increasing the value of $J$, all particles come together to yield a single wave packet, indicating the appearance of phase separation. (b) In the case of $n=0.5$, every four particles yield a wave packet in the case of the supersymmetry limit or in the case of the superconducting instability ($J=1.1$). Further increasing the value of $J$, all particles come together to yield an antiferromagnetic island.}
\end{figure*}

When $J\ge 1.2$, the structure factor $N(k)$ of the system shows the quadratic behavior near small $k$, as shown in Figs. \ref{normal_multi_Nk_rightHL80_su4tJchain}(b) and \ref{normal_multi_Nk_rightHL80_su4tJchain}(c). This indicates that the charge degree of freedom is frozen with a gap and thus the system goes into an insulating state according to Eq. (\ref{eq:gapedSUNNk}). As shown in Figs. \ref{normal_multi_Sk_rightHL80_su4tJchain}(b) and \ref{normal_multi_Sk_rightHL80_su4tJchain}(c), when $J=1.2$, the structure factor $T^{23}(k)$ has a sharp peak at $k=\pi/2$, which indicates the instability to the flavor antiferromagnetic island. In addition, the corresponding momentum distribution function is almost uniform in the whole $k$ space, as shown in Figs. \ref{normal_multi_nnk_rightHL80_su4tJchain}(b) and \ref{normal_multi_nnk_rightHL80_su4tJchain}(c). According to the uncertainty principle, the corresponding fermions of the system are localized in real space, which also indicates the system goes into an insulating state and thus is consistent with the quadratic behavior of $N(k)$ near small $k$. As shown in Fig. \ref{normal_leftright_multi_rhox_rightHL80}(b), from its corresponding density distribution in real space we can clearly see that the system goes into the phase separation \cite{PhysRevB.53.1866} area with electron density $n=1$. The flavor antiferromagnetic island with electron density $n=1$ is equivalent to the SU(4) antiferromagnetic chain forming an electron solid phase. The formation of the antiferromagnetic island indicates that there are a large number of degenerate states near the ground state, and these states can be associated with each other by translation operations. In addition, we observe that the slope of $T^{23}(k)/2$ in the case of $n=0.5$ and $J=1.2$ (labeled as ``$J=1.2$ naive") near small $k$ evidently deviates from those of other curves as shown in Fig. \ref{normal_multi_Sk_rightHL80_su4tJchain}(b). This results from the fact that the spatial range of the flavor degrees of freedom is reduced to half of the size of the system since the charge degree of freedom is frozen. In addition, this point can be clearly observed in its density profile with the feature of $n=1$ where all particles are concentrated in the right half of the region of the system. Therefore, we use the effective size of the system, $L^{\mathrm{effective}}=L/2=40$ sites, to replace the naive $L$ in the prefactor of the Fourier transformation in Eq. (\ref{eq:T23correlation}), and the corresponding data labeled as ``$J=1.2$ correct" show the same slope near small $k$ with those of other curves, as shown in Fig. \ref{normal_multi_Sk_rightHL80_su4tJchain}(b). Now it is consistent with our theory as said in Sec. \ref{sec:pheno-bosonization-analysis} that due to the flavor rotation invariance of the system, $K_{\sigma}=1$ remains unchanged for gapless flavor sectors. One can observe that this phenomenon is negligible in the high-density case (namely, the case of $n=0.9$ with $J=1.2$) since the effective size of the system is very close to the original size, as shown in Fig. \ref{normal_multi_Sk_rightHL80_su4tJchain}(c).

Now we discuss the case of $n=0.1$ with $J=1.1$. In this case, the system is close to the boundary between the molecular superfluid area and the phase separation area. As said above its flavor gap is opened. But its structure factor $N(k)$ near $k=0$ presents a dramatic peak, as shown in Fig. \ref{normal_multi_Nk_rightHL80_su4tJchain}(a). This indicates the system has a growing trend to develop a long-wavelength CDW instability, denoting the appearance of phase separation. Its density distribution in real space indicates all particles are confined to a single wave packet, as shown in Fig. \ref{normal_leftright_multi_rhox_rightHL80}(a), consistent with the feature of phase separation. One can easily observe that the value of $N(k)$ at the smallest non-vanishing value of momentum is very close to the peak point and thus remains a singularity. Note that $N(k=0)=0$ remains unchanged in all cases since the system size is finite in our calculations. Therefore, the singularity of $N(k)$ around $k=0$ can be only traced to the smallest non-vanishing value of momentum. In addition, the $K_{\rho}=5.468$ listed in Fig. \ref{normal_multi_Nk_rightHL80_su4tJchain}(a) is not so reliable due to this singularity. As shown in Fig. \ref{normal_multi_Sk_rightHL80_su4tJchain}(a), its structure factor $T^{23}(k)$ without the feature of the maximum value occurring in the flavor antiferromagnetic vector $k=\pi/2$ indicates the formation of phase separation and the formation of the antiferromagnetic island do not occur simultaneously. Its Fermi surface has been broken seriously, and the momentum distribution near $k_{F}$ becomes flat, as shown in Fig. \ref{normal_multi_nnk_rightHL80_su4tJchain}(a). 

\begin{figure}
\centering
\includegraphics[width=0.48\textwidth]{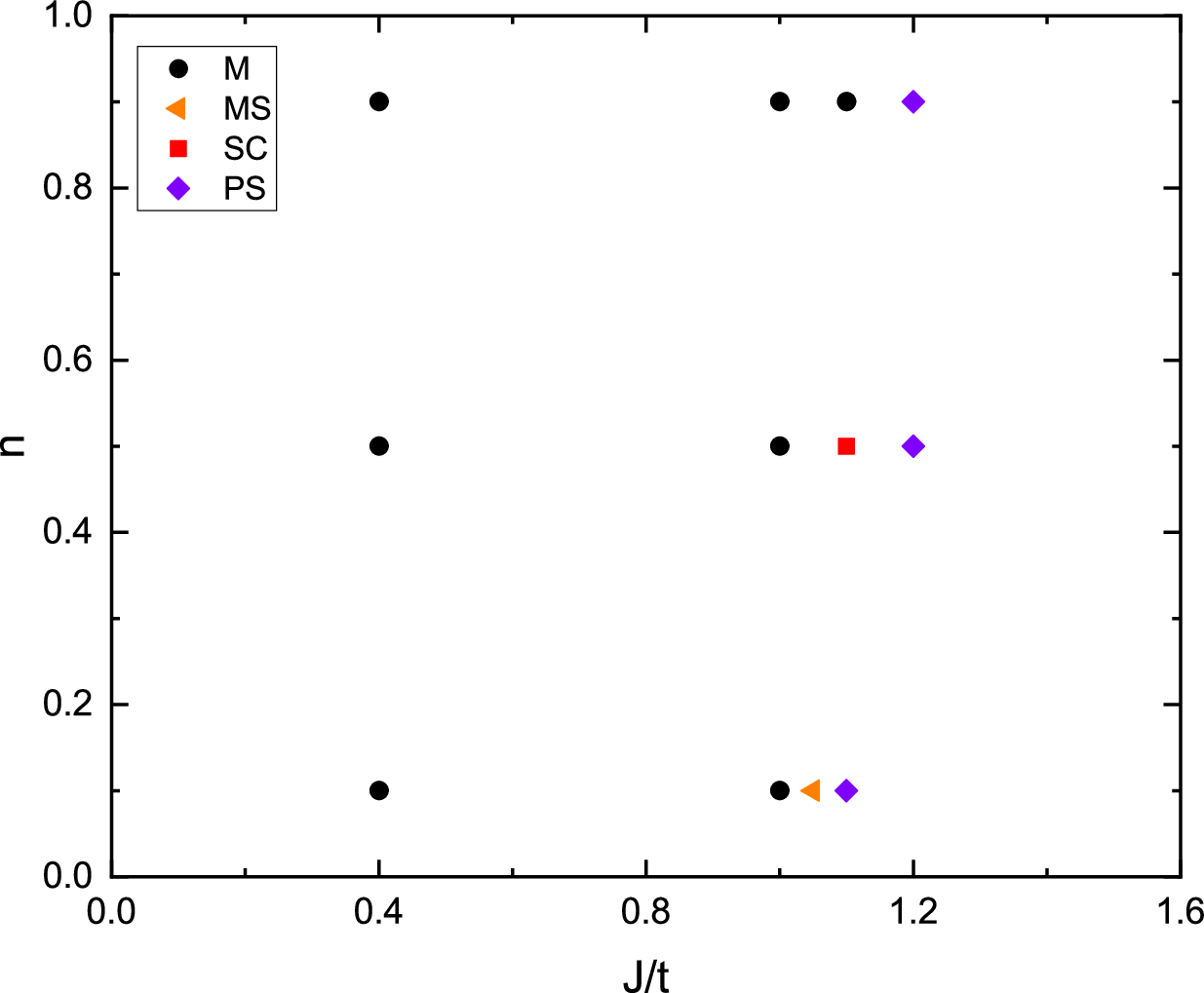}
\caption{\label{phasediagram_su4tjchain_rightHL80}Phase diagram of the SU(4) $t$-$J$ chain from DMRG, where we set $t=1$. We present four phases including the metallic phase (M) or repulsive Luttinger liquid phase, the molecular superfluid phase (MS), the superconducting phase (SC), and phase separation (PS).}
\end{figure}

We summarize our numerical results in the phase diagram shown in Fig. \ref{phasediagram_su4tjchain_rightHL80}.

\section{area of the superconducting instability}\label{sec:SC-area-without-the-flavor-gap}
\begin{figure}
\centering
\includegraphics[width=0.48\textwidth]{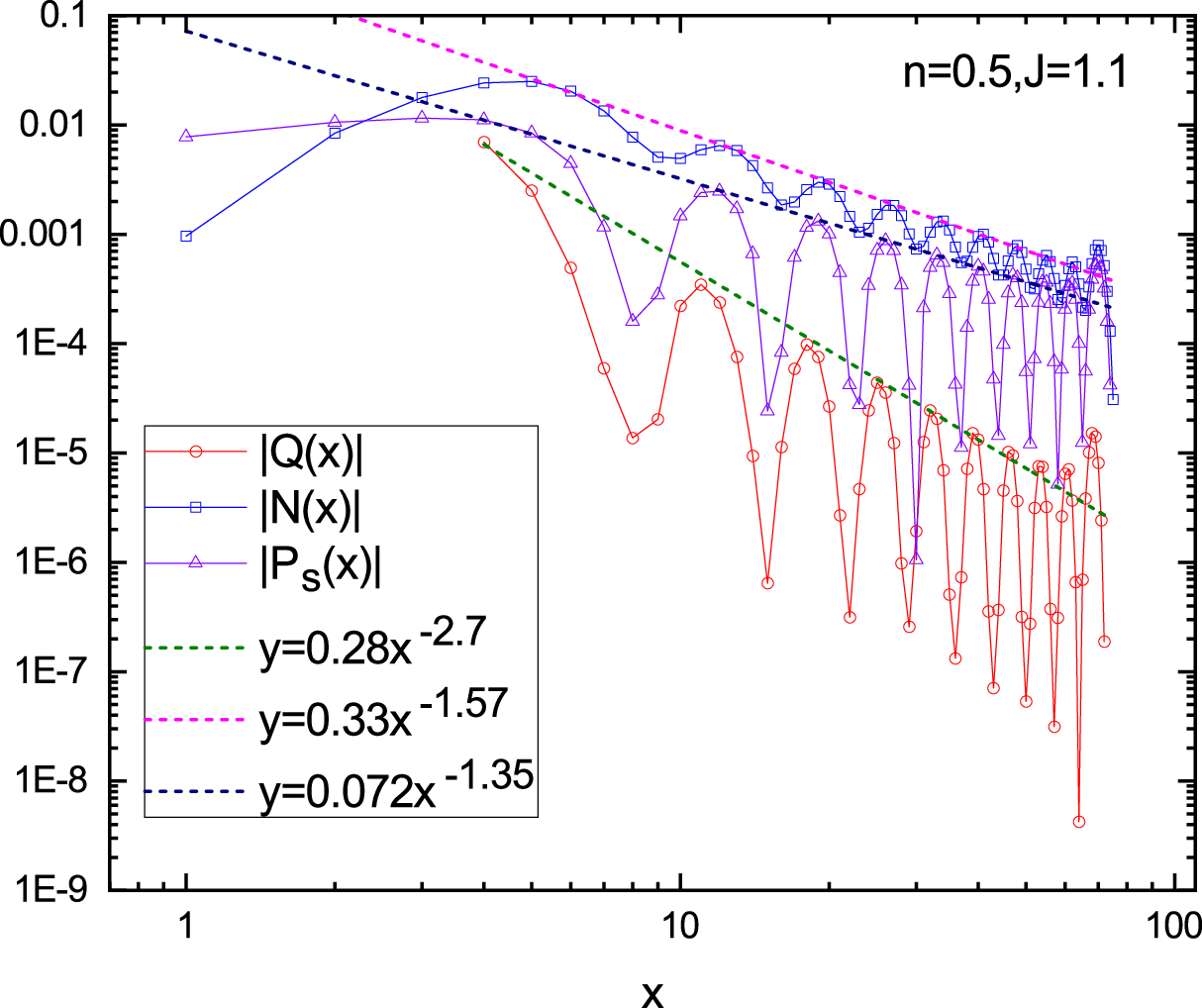}
\caption{\label{correlations_compare_rightHL80n0d5J1d1}The comparison among the power law decay behaviors (on a double-logarithmic scale) of several correlation functions including the SU(4) singlet correlation function $Q(x)=Q_{ij}$, the density-density correlation function $N(x)=N_{ij}$, and the pairing correlation function $P_{s}(x)=P_{s,ij}$ corresponding to the pairing field $\vec{\Delta}_{i j}$. Here $x=j-i$, ($j>i$). The system size is given by $L=80$ sites. The straight dot lines are fitting curves to extract the exponents.}
\end{figure}

For the case of $n=0.5$ with $J=1.1$, as analyzed in Sec. \ref{sec:structure-factors}, there is the superconducting instability rather than the molecular superfluid instability. To verify our theoretical analysis, we calculate the correlation function $Q(x)=Q_{ij}$ of the SU(4) singlet, the density correlation function $N(x)=N_{ij}$, and the correlation function $P_{s}(x)=P_{s,ij}$ of the pairing field $\overrightarrow{\Delta}_{i j}$. Here $x=j-i$, ($j>i$). Their decay behaviors in real space are shown in Fig. \ref{correlations_compare_rightHL80n0d5J1d1} where for exhibiting the power law behaviors of the correlation functions we adopt a double-logarithmic scale. We can see that these correlation functions all have the power-law behaviors in real space, and it is evident that the pairing correlation function $P_{s}(x)$ decays slower than the other two correlation functions. The exponents of these three kinds of correlation functions are extracted by fitting curves. The extracted exponent of the pairing correlation function is evidently much smaller than two and thus indicates its susceptibility is divergent. In contrast, the extracted exponent of the correlation function of the SU(4) singlet is unambiguously much bigger than two and thus indicates its susceptibility is not divergent. This qualitative conclusion is indeed consistent with our theoretical analysis of the phenomenological bosonization results. The theoretical exponents of the leading term of the pairing correlation function and that of the correlation function of the SU(4) singlet are given by $(K_{\rho}^{-1}-1)/2+2=1.930$ and $2+2K_{\rho}^{-1}=3.721$, respectively, using the extracted $K_{\rho}=1.162$. In addition, the theoretical exponent of the leading term of the density correlation function is given by $(K_{\rho}-1)/2+2=2.081$. These three theoretical exponents are all bigger than the corresponding extracted exponents shown in Fig. \ref{correlations_compare_rightHL80n0d5J1d1}. We propose the contribution from the subleading terms of their correlation function is responsible for this behavior and this behavior may reduce with increasing the size of the system. Especially, one can find that, for the density correlation function, the numerically extracted exponent $1.57<2$ is qualitatively different from the theoretical exponent $2.081>2$. We propose that the value of the theoretical exponent $2.081$ is so close to the divergent value $2$ that any small corrections to the leading term may lead to the divergence of its susceptibility. In fact, as shown in Fig. \ref{normal_multi_Nk_rightHL80_su4tJchain}(b), its $2k_{F}$ peak of $N(k)$ is not sharp and is just a broad maximum, signaling the tendency to the $2k_{F}$ CDW is not strong. Therefore, the pairing field $\overrightarrow{\Delta}_{i j}$ is dominant in this case.

We can directly observe the Friedel oscillations \cite{Friedel:1952vf,Mahan1981} in density from our numerical results due to the open boundary condition. This phenomenon can show the feature of the pairing in real space. Before discussing it, let's use the case of the low density $n=0.1$ to demonstrate the Friedel oscillations. In this case when $J=0.4$, the repulsive Luttinger liquid shows clearly density oscillations with eight packets, indicating every particle yields a wave packet, as shown in Fig. \ref{normal_leftright_multi_rhox_rightHL80}(a). When the value of $J$ increases to $J=1$ (the supersymmetry limit), the number of oscillations reduces to two, indicating every four particles with flavors different from each other yield a wave packet, but these four particles do not form a bound state since no actual binding energy is involved \cite{Schlottmann_1993}. With the value of $J$ increasing to $J=1.05$, the number of oscillations remains two, but the wave packet yielded by four particles is more pronounced, which is consistent with the scenario that the $2k_{F}$ CDW instability coexists with the molecular superfluid instability. This can be understood by the scenario that the behavior of the bound SU(4) singlet states resembles that of the hard-core bosons due to the occupancy constraint of the SU(4) $t$-$J$ model. In the case of the intermediate density $n=0.5$, the oscillations are not evident in the repulsive Luttinger liquid when $J=0.4$ since the average distance between particles is so small, as shown in Fig. \ref{normal_leftright_multi_rhox_rightHL80}(b). But we can clearly observe that oscillations occur in the superconducting phase when $J=1.1$, indicating every wave packet is also yielded by four particles with flavors different from each other. This pairing is special since a six-component pairing consists of four particles. We can still think the behavior of the six-component pairing of $\overrightarrow{\Delta}_{i j}$ resembles that of the hard-core boson. Therefore, the behavior of the pairing of $\overrightarrow{\Delta}_{i j}$ in real space is almost like that of the SU(4) singlet.

\section{Discussion and conclusion}\label{sec:Discussionandconclusion}
The term $(1-2/N)K_{\sigma}^{-1}$ in exponents of the pairing correlation function (Eq. (\ref{eq:antisym-pairing-correl})) makes the difference between the SU(2) case and the SU($N$) case with $N>2$. When $N=2$, this term vanishes and all the exponents of the pairing correlation function are the same as those of the correlation function of the SU(2) singlet. This is natural since these two correlation functions are the same thing in the SU(2) case. However, when $N>2$ this term becomes nonvanishing. It becomes infinite when the flavor gap is opened, which leads to the pairing correlation function being exponentially suppressed and thus means the flavor gap and the pairing are separated, in contrast to the SU(2) case.

The area of the parameters of the molecular superfluid instability is small and only distributes in the low density, according to Fig. \ref{normal_multi_Sk_rightHL80_su4tJchain} and our numerical results. To study the molecular superfluid instability in low density and obtain the definite phase boundaries of the whole phase diagram, we need to consider the larger size of the system and perform finite-size extrapolations to obtain the value of the flavor gap and other quantities. However, the larger size is not practical in numerical calculations due to the rapidly increasing calculation time. We leave these studies to future work.

In conclusion, we demonstrate the SU($N$) singlet instability cannot coexist with the pairing instability in the one-dimensional models with SU($N$) symmetry when $N>2$ by our theoretical analysis of the phenomenological bosonization results. We use the DMRG method to study the SU(4) $t$-$J$ chain. Our numerical results show there are two phases in the attractive Luttinger liquid region. The first is the molecular superfluid phase (the SU(4) singlet instability) where the CDW instability is more dominant than the molecular superfluid instability in the case of $n=0.1$ with $J=1.05$. The second is the superconducting phase (the six-component pairing instability). Our numerical results indeed verify the theoretical analysis of bosonization results.

\begin{acknowledgments}
We are grateful for the invaluable discussions with T. K. Lee. This work was supported by the National Key Research and Development Program of China Grant No. 2022YFA1404204, the National Natural Science Foundation of China (Grants No. 11625416 and No. 12274086).
\end{acknowledgments}

\appendix
\section{Interaction term of the SU($N$) $t$-$J$ model, and time-dependent correlation functions of SU($N$) models}\label{sec:time-dependent-correl}
The interaction term of the SU($N$) $t$-$J$ model in Eq. (\ref{eq:SUNt-Jmodelv1}), by using the SU($N$) generators, can be rewritten as
\begin{flalign}
\frac{J}{N}\sum_{i}\sum_{a=1}^{N^{2}-1} \hat{T}_{i}^{a} \hat{T}_{i+1}^{a},
\end{flalign}
where we neglect the particle number terms. Here $\hat{T}_{i}^{a}=\sum_{\alpha\beta}c_{i, \alpha}^{\dagger} \Gamma_{\alpha \beta}^{a} c_{i, \beta}$, with
$\Gamma^{a}$ being the generator of SU($N$) group and satisfying the relation $\mathrm{Tr}\left(\Gamma^{a}\Gamma^{b}\right)=N\delta_{ab}$. This interaction term by using the Fierz identity can be rewritten as
\begin{flalign}
\sum_{a=1}^{N^{2}-1} \hat{T}_{i}^{a} \hat{T}_{j}^{a}=-\frac{N+1}{N}\left(\overrightarrow{\Delta}_{i j}\right)^{\dagger} \cdot \overrightarrow{\Delta}_{i j}+\frac{N-1}{N}\left(\Delta_{i
j}^{-}\right)^{\dagger} \cdot \Delta_{i j}^{-},
\end{flalign}
where the two pairing fields can be expressed as:
\begin{flalign}
&\overrightarrow{\Delta}_{i j}=\sum_{m=1}^{N(N-1)/2}\left(\sum_{\alpha\beta}c_{i\alpha}\Gamma_{\alpha\beta}^{n_{m}}c_{j\beta}\right)\hat{\mathbf{e}}_{m},
                        \nonumber\\
&\Delta_{i j}^{-}=\sum_{m=1}^{N(N+1)/2}\left(\sum_{\alpha\beta}c_{i\alpha}\Gamma_{\alpha\beta}^{p_{m}}c_{j\beta}\right)\hat{\mathbf{e}}_{m},
\end{flalign}
Where we use the superscript $n_{m}$ to denote the antisymmetric generators of SU($N$) group, and the superscript $p_{m}$ to denote the generators of SU($N$) group including the symmetric elements and the identity element. $\hat{\mathbf{e}}_{m}$ represents an orthogonal normalized vector basis. We only consider the lower-level pairing field $\overrightarrow{\Delta}_{i j}$ due to $J>0$. We only consider one of the $N(N-1)/2$ components of $\overrightarrow{\Delta}_{i j}$ due to the symmetry between them, such as
\begin{flalign}
\Delta_{i}^{s\dagger}=\left(c_{i, 1}^{\dagger} c_{i+1,2}^{\dagger}-c_{i, 2}^{\dagger} c_{i+1,1}^{\dagger}\right) / \sqrt{2}.
\end{flalign}

Now we give time-dependent correlation functions of SU($N$) models at zero temperature for the case of the gapless regime in the charge sector and the flavor sectors. The total density correlation function is given by
\begin{flalign}\label{eq:pheno-boson-totdensityv2}
&\langle \rho(x,\tau)\rho(0,0) \rangle=n^{2}+\frac{NK_{\rho}}{2\pi^{2}}\frac{y^{2}_{\rho}-x^{2}}{(x^{2}+y^{2}_{\rho})^{2}}
                                       \nonumber\\
&+\sum_{p=1}^{\infty}A_{p+1}\cos(2pk_{F}x)\left(\frac{\alpha}{r_{\rho}}\right)^{2p^{2}K_{\rho}/N}\left(\frac{\alpha}{r_{\sigma}}\right)^{2p^{2}(1-1/N)K_{\sigma}},
\end{flalign}
where $y_{\rho}=u_{\rho}\tau+\alpha \mathrm{Sign}(\tau)$, and $r_{\nu}=\sqrt{x^2+(u_{\nu}\tau)^2}$ with $\nu=\rho$, $\sigma$. In addition, $u_{\rho}$ and $u_{\sigma}$ are velocities of the excitation in the charge sector and the flavor sector, respectively. The flavor-flavor correlation function is given by
\begin{flalign}\label{eq:pheno-boson-flavorcorrelv2}
&\langle [\rho_{1}(x,\tau)-\rho_{2}(x,\tau)][\rho_{1}(0,0)-\rho_{2}(0,0)]\rangle
           \nonumber\\
&=\frac{K_{\sigma}}{\pi^{2}}\frac{y_{\sigma}^{2}-x^{2}}{(x^{2}+y^{2}_{\sigma})^{2}}
                 \nonumber\\
&+\sum_{p=1}^{\infty} B_{p+1} \cos(2pk_{F}x)\left(\frac{\alpha}{r_{\rho}}\right)^{2p^{2}K_{\rho}/N}\left(\frac{\alpha}{r_{\sigma}}\right)^{2p^{2}(1-1/N)K_{\sigma}},
\end{flalign}
where $y_{\sigma}=u_{\sigma}\tau+\alpha \mathrm{Sign}(\tau)$. The correlation function of the SU($N$) singlet state is given by
\begin{flalign}\label{eq:SU(N)singletcorrelbosonevenv2}
&\langle\widetilde{\mathcal{M}}^{\dagger}(x,\tau)\widetilde{\mathcal{M}}(0,0)\rangle
                         \nonumber\\
&=\sum_{p=0}^{\infty}C_{p}\left(\frac{\alpha}{r_{\sigma}}\right)^{(2p+1)^{2}NK_{\sigma}/2}\left(\frac{\alpha}{r_{\rho}}\right)^{NK_{\rho}^{-1}/2}+\cdots.
\end{flalign}
Note that $N$ in Eq. (\ref{eq:SU(N)singletcorrelbosonevenv2}) is restricted to even numbers. The case of odd numbers for $N$ is given by
\begin{flalign}\label{eq:SU(N)singletcorrelbosonoddv2}
&\langle\widetilde{\mathcal{M}}^{\dagger}(x)\widetilde{\mathcal{M}}(0)\rangle=\sum_{p}C^{\prime}_{p}e^{i(2p+1)[k_{F}x-\mathrm{Arg}(y_{\rho}+ix)]}
               \nonumber\\
&\left(\frac{\alpha}{r_{\rho}}\right)^{(2p+1)^{2}K_{\rho}/(2N)+NK_{\rho}^{-1}/2}\left(\frac{\alpha}{r_{\sigma}}\right)^{(2p+1)^{2}(N-1/N)K_{\sigma}/2}
               \nonumber\\
&+\cdots.
\end{flalign}
The correlation function of the flavor-antisymmetric pairing is given by
\begin{flalign}
&\langle\Delta^{s\dagger}(x,\tau)\Delta^{s}(0,0)\rangle
                                \nonumber\\
&=\sum_{p=0}^{\infty}D_{p}\left(\frac{\alpha}{r_{\rho}}\right)^{2/(NK_{\rho})}\left(\frac{\alpha}{r_{\sigma}}\right)^{(1-2/N)K_{\sigma}^{-1}+(2p+1)^{2}K_{\sigma}}
                                \nonumber\\
&+\cdots.
\end{flalign}

\bibliography{SU4tJchain}

\end{document}